\documentclass[reprint,amsmath,amssymb,aps]{revtex4-1}
\usepackage{graphicx}
\usepackage{bm}
\usepackage{xcolor}
\usepackage{float}

{
\begin{document}
\title{ Emergent Dynamics and Spatio Temporal Patterns on Multiplex Neuronal Networks }%
\author{Umesh Kumar Verma and G. Ambika}
\affiliation{Department of Physics, Indian Institute of Science Education and Research(IISER) Tirupati, Tirupati, 517507, India}

\begin{abstract}
We present a study on the emergence of a variety of spatio temporal patterns among neurons that are connected in a multiplex framework, with neurons on two layers with different functional couplings. With the Hindmarsh-Rose model for the dynamics of single neurons, we analyze the possible patterns of dynamics in each layer separately and report emergent patterns of activity like in-phase synchronized oscillations and amplitude death for excitatory coupling and anti-phase mixed-mode oscillations in multi-clusters with phase regularities when the connections are inhibitory. When they are multiplexed, with neurons of one layer coupled with excitatory synaptic coupling and neurons of the other layer coupled with inhibitory synaptic coupling, we observe transfer or selection of interesting patterns of collective behavior between the layers. While the revival of oscillations occurs in the layer with excitatory coupling, the transition from anti-phase to in-phase and vice versa is observed in the other layer with inhibitory synaptic coupling. We also discuss how the selection of these spatio temporal patterns can be controlled by tuning the intralayer or interlayer coupling strengths or increasing the range of non-local coupling.  With one layer having electrical coupling while the other synaptic coupling of excitatory(inhibitory)type, we find in-phase(anti-phase) synchronized patterns of activity among neurons in both layers.

 
\end{abstract}
\maketitle
\section{Introduction}

The complexity underlying the patterns of dynamical behavior in the brain is a fascinating and challenging research area in recent times~\cite{Sporns}. The complexity arises not only from the large number of neurons involved but also from the variety and plasticity of connections or interactions among them during any type of neuronal or cognitive activity ~\cite{Pereda, Ashwin}. The interactions can be electrical via gap junction and excitatory or inhibitory interaction via chemical synapses. The collective behavior or synchronization among a large number of neurons is essential for various neurobiological processes, which mostly appear due to the inter neuronal synaptic interactions~\cite{Pikovsky01}. Also, various brain disorders such as Alzheimer's disease, schizophrenia, Parkinson's disease, and epilepsy have been linked to the abnormal patterns of synchronization among the neurons~\cite{Uhlhaas, Jalili, Knyazeva}. The nature of the collective dynamics can have different forms of oscillatory patterns that include in-phase oscillations, anti-phase oscillations, multi-cluster oscillations etc~\cite{Jalan, Pournaki}. In addition, coupled neurons also show quiescent states due to suppression of activity or amplitude death (AD)~\cite{AD1}.

We find the multiplex framework is ideal for describing the collective dynamics of neurons since an assembly of neurons can have excitatory or inhibitory types of electrical or chemical synaptic interactions~\cite{Boccaletti, Umesh1}. Then analysis can be done with the same set of neurons distributed in different layers, based on the nature of interactions among them. In the present study, we consider the framework of multiplex networks to study the activity patterns that can emerge or get selected when neurons in one layer interact with each other through excitatory synaptic couplings and neurons in the other layer interact with each other through inhibitory synaptic couplings. Equally interesting and realistic is the case where one layer of neurons interact electrically while in the other layer, the interaction is synaptic or chemical of excitatory or inhibitory type. We begin by studying the patterns of collective dynamics in each layer separately and observe how excitatory synaptic coupling induces completely synchronized oscillations and amplitude death, while inhibitory synaptic coupling induces anti-phase synchronized oscillations for local connections and multi-cluster oscillations with relative phase ordering for nonlocal connections. In this context, we note that anti-phase synchronization is observed in neuronal networks in human and animal brains~\cite{Ueda, Ohta, Iglesia}, climactic networks~\cite{Saenko, Hinnov}, food web~\cite{Vandermeer}, and lasers~\cite{Wiesenfeld}. We note in multiplex neuronal networks with attractive and repulsive interactions, anti-phase synchronization is reported recently ~\cite{Chowdhury} and chimera states are found to occur in multilayer networks of neurons ~\cite{Majhi0,Majhi1,Majhi2}.

When both layers are multiplexed, we find transfer or selection of activity patterns across the layers, with the revival of oscillations from amplitude death state in the first layer and a transition from anti-phase to in-phase in the second layer. Depending on the strength of intralayer coupling, activity patterns corresponding to the stronger interaction get selected and stabilized across the neurons in both layers. When one layer has electrical coupling and the other layer with synaptic coupling, in-phase or anti-phase oscillations are induced depending on whether synaptic coupling is excitatory or inhibitory. These activity patterns have rhythmic dynamics with mixed-mode oscillations(MMO), which are complex periodic forms of activity. We note such MMOs are experimentally observed and analyzed in neurophysiological studies~\cite{Desroches,Negro, Ghosh1}. We study the transitions between such patterns of activity and how the relevant parameters can be tuned for a specific pattern to get selected across the layers.

\section{Multiplex neuronal networks}

We consider a multiplex network of neurons with two layers, each of them consisting of an ensemble of $N$ Hindmarsh-Rose (HR) neurons coupled on a regular ring network. We take the neurons in the first layer (L1) to be interacting with each other with excitatory synaptic coupling and those in the second layer (L2) interacting through an inhibitory synaptic coupling. The neurons in L1 interact with neurons in L2 with multiplex like $i$ to $i$ coupling via feedback. The dynamics of the multiplex network of neurons is thus modelled as,

\begin{eqnarray}
\dot{x}_{i,1}&=&B_{i,1}+ \frac{\lambda_1}{2p_1}(V_s-x_{i,1})\sum_{k=i-p_1}^{i+p_1}\Gamma (x_{k,1})+\epsilon  x_{i,2} \nonumber\\
\dot{y}_{i,1}&=&(a+\alpha)x_{i,1}^2-y_{i,1} \nonumber\\
\dot{z}_{i,1}&=&c(bx_{i,1}-z_{i,1}+e) \nonumber\\
\dot{x}_{i,2}&=&B_{i,2}-\frac{\lambda_2}{2p_2}(V_s-x_{i,2})\sum_{k=i-p_2}^{i+p_2}\Gamma (x_{k,2})+\epsilon  x_{i,1} \nonumber\\
\dot{y}_{i,2}&=&(a+\alpha)x_{i,2}^2-y_{i,2} \nonumber\\
\dot{z}_{i,2}&=&c(bx_{i,2}-z_{i,2}+e)
\label{eq10}
\end{eqnarray}

where $B_{i,j}=ax_{i,j}^2-x_{i,j}^3-y_{i,j}-z_{i,j}$, $i=1,2,...,N$ and $j=1,2$~\cite{Majhi0}. The variable $x_{i,j}$ represents the action potential and the variable $y_{i,j}$ and $z_{i,j}$ represent the transport of ions across the membrane through fast and slow channels respectively. The function $\Gamma (x_{i,j})=1/\{1+\exp[-\beta(x_{i,j}-\phi_s)]\}$  is the sigmoidal chemical synaptic coupling function with $V_s$ as reversal potential. Here we take the reversal potential $V_s = 2$ such that $V_s > x_i(t)$ can be satisfied. We choose the synaptic threshold $\phi_s =-0.25$ and $\beta=10$ in the sigmoidal function. Also, $p_1$ and $p_2$ take care of the range of interactions, whether it is local or nonlocal, with $p_{1,2}=1$ being local. The other system parameters are $a = 2.8$, $\alpha=1.6$, $b=9$, $c=0.001$, and $e=5$ such that the individual HR neurons shows regular square-wave bursting dynamics. In the present work, the emergent dynamics of Hindmarsh-Rose (HR)neurons are studied by solving Eq.~\ref{eq10}, using fourth-order Runge-Kutta method, with initial conditions are chosen randomly between $-1$ to $1$, for various cases as presented below.

\begin{figure}
\includegraphics[width=0.45\textwidth]{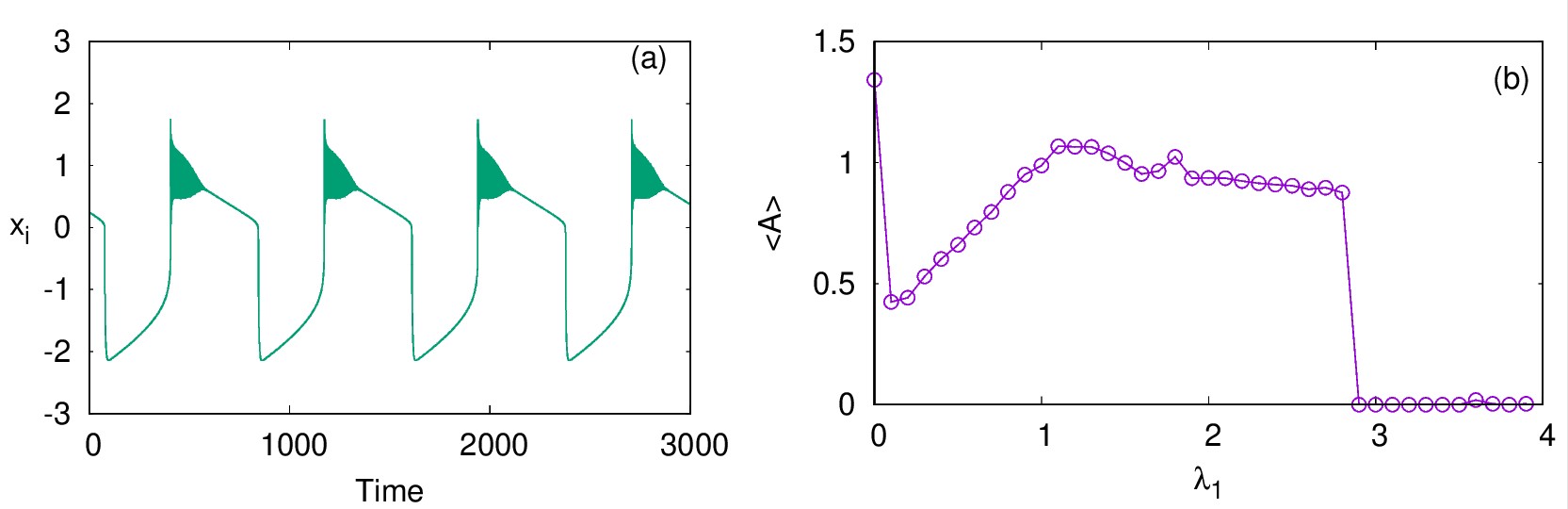}
\caption{(a)Time series showing synchronized bursts of action potentials from all the neurons on layer L1, coupled with excitatory synaptic coupling for coupling strength $\lambda_1=1.5$, (b) Average amplitude of oscillations for increasing coupling strength $\lambda_1$. The transition from oscillatory state to amplitude death state occurs at $\lambda_1=2.9$.  Here $p_1=1$, $\epsilon=0$ and $N=50$.}
\label{f1}   
\end{figure} 

\subsection{Spatio temporal patterns on single layer}

We begin by considering the emergent dynamics or patterns of activity that can develop in each layer in the absence of multiplexing with $\epsilon=0$ and number of neurons $N=50$ in Eq.~\ref{eq10}.  

In layer L1 with excitatory synaptic coupling among neurons, we observe that, for sufficient strength of synaptic coupling, they settle to a completely synchronized oscillatory state, which is shown in Fig.~\ref{f1}(a) at $\lambda_1=1.5$. However, the nature of oscillations is changed from intrinsic bursts to varied forms like bursts of decreasing amplitudes and broad spikes as $\lambda_1$ is increased.  With stronger coupling, at $\lambda_1=2.9$, these spikes are suppressed, and the layer goes to amplitude death(AD). We note amplitude death phenomenon has been reported earlier in globally coupled HR neurons~\cite{Prasad}. Here we find that amplitude death can occur for all values of $p_1$, local, nonlocal and global, with sufficient strength of coupling. To detect the transition to AD, we compute the average amplitude of the spikes of all the neurons using ~\cite{Umesh} 

\begin{equation}
<A>=(\sum_{i=1}^N \langle x_{i,max}\rangle_t )/N
\end{equation}
This is plotted is in Fig.~\ref{f1}(b) for $p_1=1$ with increasing $\lambda_1$. We find the average amplitude increases with $\lambda_1$ initially, reaches a maximum, and then decreases. At $\lambda_1=2.9$, there is a sudden transition to AD. The nature of the burst patterns in these regions differs as spikes of decreasing amplitude in each burst that change to square bursts before reaching AD. We repeat the study by increasing N to 100 and 500 and find qualitatively similar results.

\begin{figure}
\includegraphics[width=0.45\textwidth]{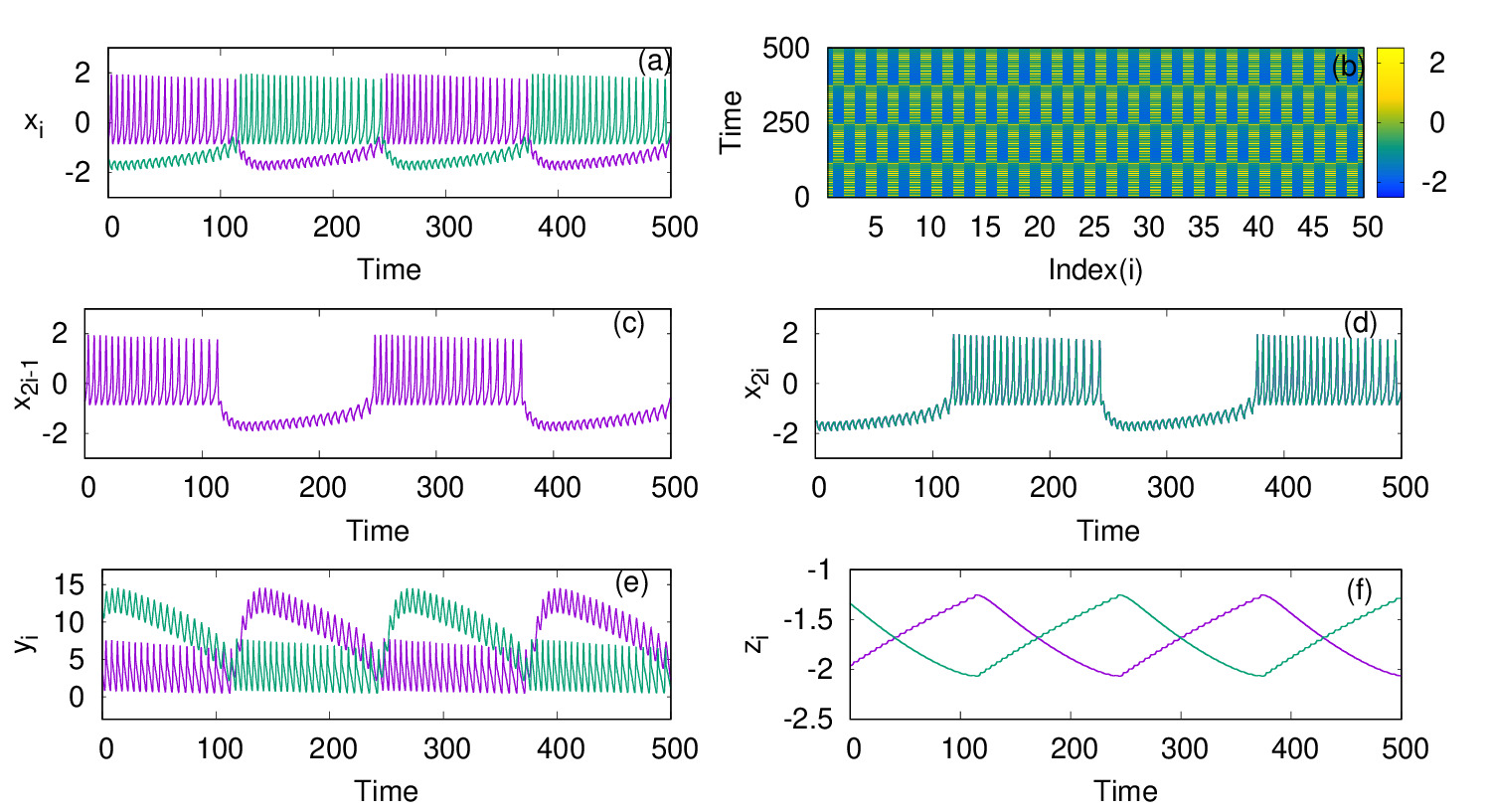}
\caption{(a)Mixed-mode oscillations of action potentials and (b)spatio-temporal plot of neurons coupled with inhibitory synaptic coupling on layer L2 at coupling strength $\lambda_2=1$. (c)time series of all the odd number nodes $x_{2i-1}$, showing complete synchronized oscillations, for $i =1; 2:::n/2 $ and (d)time series of all the even number nodes $x_{2i}$, where $i = 1; 2:::n/2$ that indicate completely synchronized oscillations among them. Time series of the other variables $y_i$ and $z_i$ are plotted in (e) and (f) respectively.  Here other parameters are kept as $p_2=1$, $\epsilon=0$, and $N=50$. The anti-phase nature of the oscillations in adjacent nodes and in-phase nature in alternate nodes separates the network into two clusters. The color bar in Fig. 2(b) (also in spatio-temporal plots in later figures) indicates the values of action potential $x_i$.}
\label{f2}   
\end{figure} 

For the dynamics on the second layer L2 with inhibitory synaptic coupling among neurons, we first study the case when $p_2=1$, i.e., the system has only local interactions. We find that the emergent dynamics in this case shows anti-phase synchronized oscillations, which is clear from the time series and spatio-temporal plots shown for coupling strength $\lambda_2=1$, in Fig.~\ref{f2}(a,b). First we note the nature of dynamics is changed from intrinsic bursts in this case also, revealing mixed-mode oscillations(MMO). 
Moreover, we find the neurons in one cluster, say at all even number sites, are all synchronized completely but are in anti-phase with those in the other cluster, at odd number sites. This is made more explicit by plotting the time series of all odd number of neurons and even number of neurons separately in Fig.~\ref{f2}(c)and(d), that display the pattern of anti-phase synchronized oscillations among the adjacent neurons.We also show the time series of the other two variables $y_i$ and $z_i$ in  Fig.~\ref{f2} (e) and (f) respectively. 

For a detailed characterization of the observed phase order in temporal dynamics, we calculate the phase of each neuron from its time series, $x_i$. We note the time $T_k^i$, ($k=1,2,...$) at which  $x_i$, crosses the chosen threshold value, and then we calculate the phase of the $i^{th}$ neuron using the following equation~\cite{Pikovsky01}:

\begin{figure}
\includegraphics[width=0.45\textwidth]{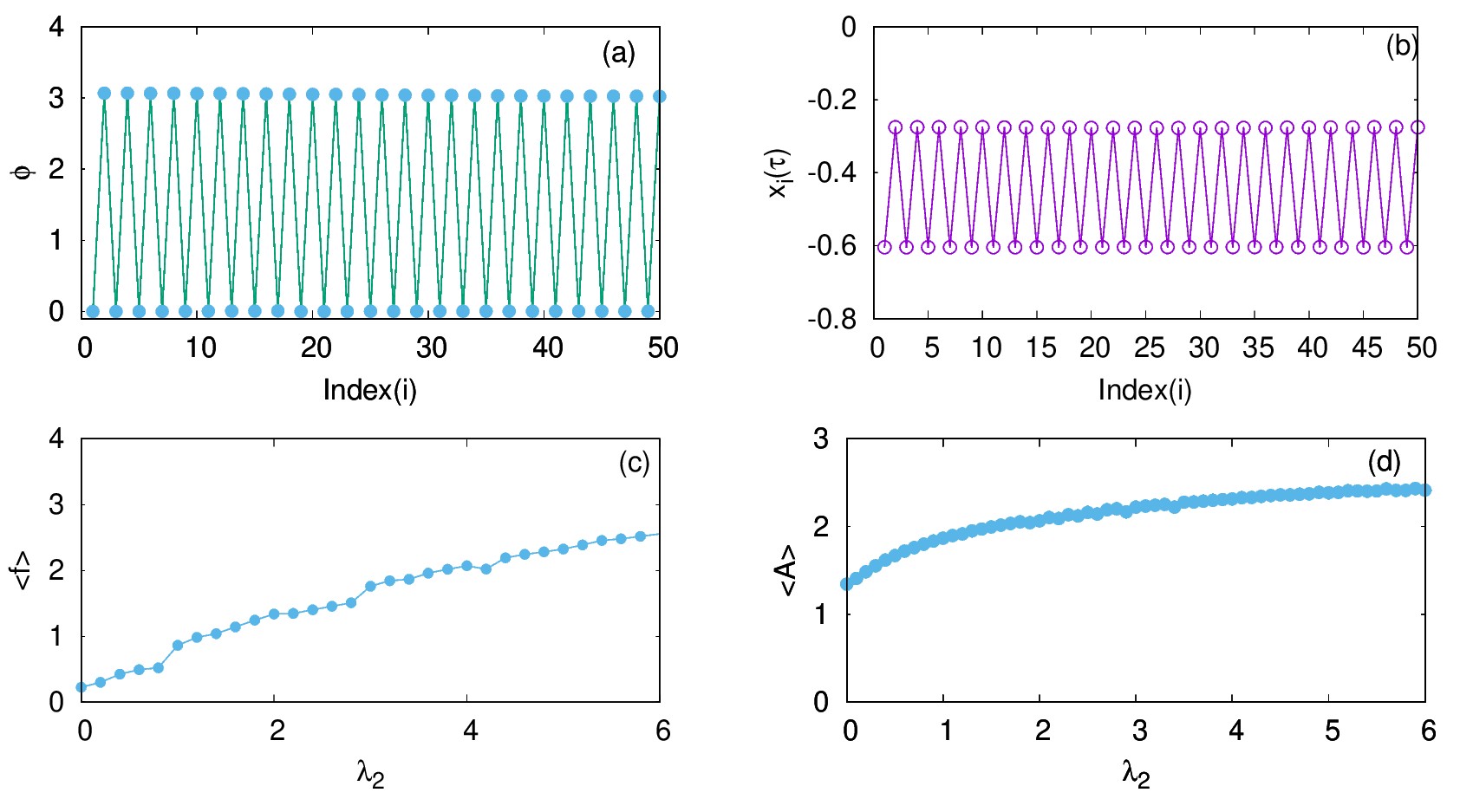}
\caption{Phase of each neuron in layer L2, coupled with inhibitory synaptic coupling calculated relative to that of its first neuron for coupling strength $\lambda_2=1$ and $p_2=1$. (b) snapshot of $x_i$ at a given time $\tau$, shows that every odd neuron is in anti-phase with every even neuron. (c) average frequency and (d) average amplitude of the large amplitude oscillations of the mixed-mode oscillations for increasing coupling strength $\lambda_2$.}  
\label{f3} 
\end{figure}

\begin{equation}
\phi_i(t)=2\pi\frac{t-T_k^i}{T_{k+1}^i-T_k^i}, \quad T_k^i\leq  t\leq T_{k+1}^i,
\label{eq2}
\end{equation}
where $i=1,2,...,N$. In Fig.~\ref{f3}(a), the phase of each neuron calculated  relative to that of the first neuron in plotted. It is clear that every odd neuron is in anti-phase with every even neuron. The snapshot of $x_i$ at a given time $\tau$ is shown in Fig.~\ref{f3}(b), that further confirms the  anti-phase pattern of the mixed mode oscillations. This is induced by the range (nearest neighbour) and the nature(inhibitory)of the coupling chosen in this context. Thus, the neurons in effect form two clusters such that synchronized oscillations in one cluster are anti-phase with that in the other cluster. Further, we calculate the spike frequency of the large amplitude oscillations of $i^{th}$ neuron as ~\cite{Mozumdar}:

\begin{equation}
f_i=\frac{2\pi}{K_i}\sum_{k=1}^{K_i}\frac{1}{t_{k+1}^i-t_k^i},
\end{equation}
where $K_i$ refers to the number of spikes for the $i^{th}$ neuron in each burst and $t_k^i$ corresponds to time of maximum of $k^{th}$ spike. Then the average frequency obtained from this, is plotted in Fig.~\ref{f3}(c) with increasing coupling strength $\lambda_2$. Here we can see that the average frequency increases with increasing $\lambda_2$. We also show how the average amplitude $<A>$ of coupled neurons increases with $\lambda_2$, for the range considered as shown in Fig.~\ref{f3}(d). Both the frequency and amplitude calculated here relate to the large amplitude spikes of the mixed-mode oscillatory states of the neurons. We note such activity patterns of synchronized oscillations with amplification are reported in multiplex networks in a different context~\cite{Njougouo}.

\begin{figure}
\includegraphics[width=0.45\textwidth]{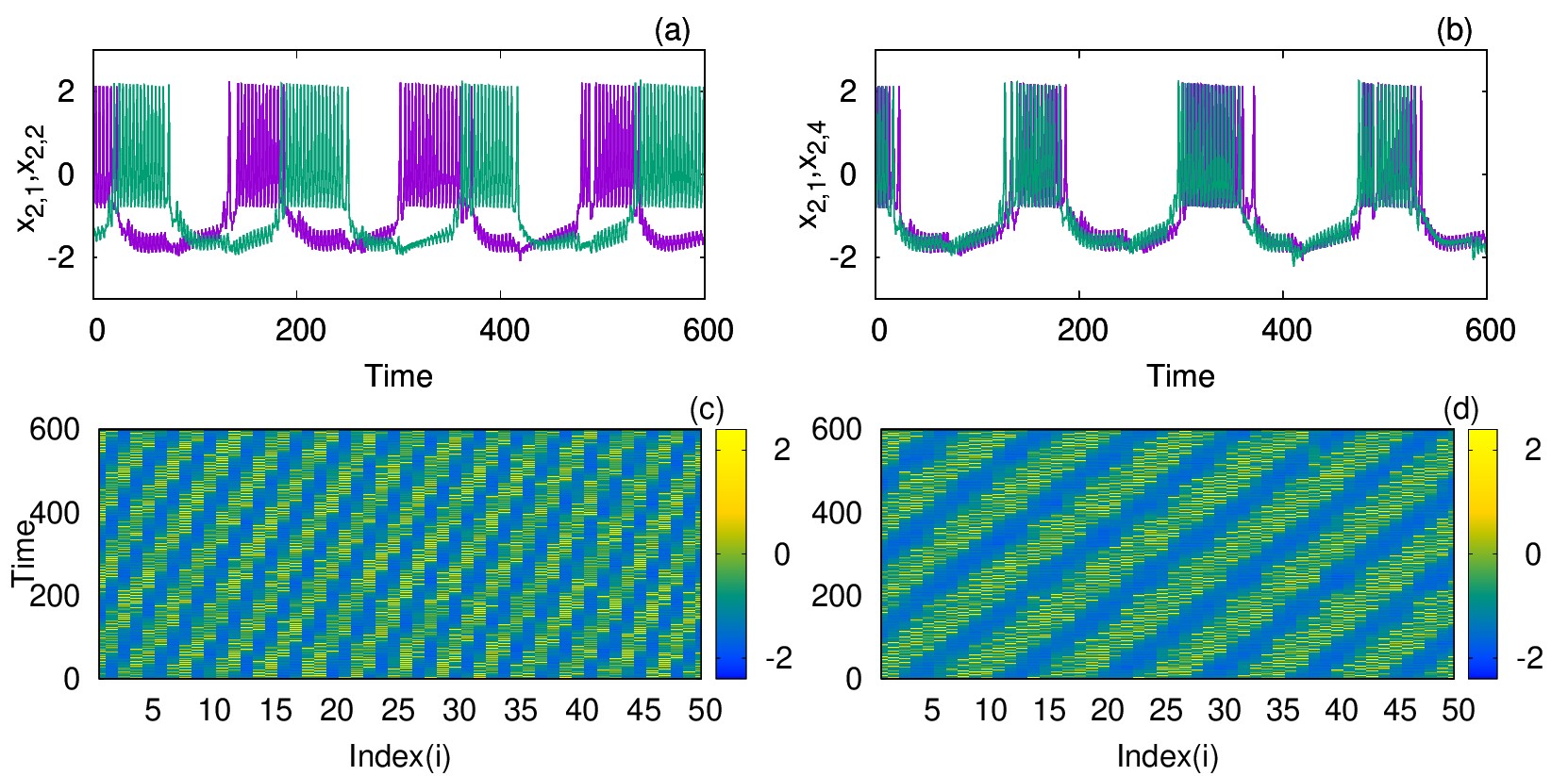}
\caption{Time series of action potentials from (a) nodes 1 and 2  and (b) nodes 1 and 4 in layer L2 coupled with inhibitory synaptic coupling for $p_2=2$. We find nodes $i$ and $i+3$ show inphase synchronized  oscillations with small phase shift. The spatio-temporal plot of neurons showing traveling wave like patterns (b1) for $p_2=2$ (b2) for $p_2=5$. Here $\lambda=3$ and $N=50$. }
\label{f4}   
\end{figure}

As the range of coupling increases or the coupling becomes nonlocal, we observe traveling wave-like patterns. In Fig.~\ref{f4}(a), we plot the time series of the action potential from node 1 and node 2 and in  Fig.~\ref{f4}(b) that from node 1 and node 4. It is clear that node 1 and 4 are almost synchronized but with a small phase shift. We find this shift in the phase depends on the coupling strength and the range of coupling. The spatio-temporal plot in this case shows travelling wave like patterns, as shown in Fig.~\ref{f4}(c,d), for $p_2=2$ and $p_2=5$, and $\lambda_2=3$. For larger sizes of networks also, we find qualitatively similar emergent dynamics.

\begin{figure}
\includegraphics[width=0.45\textwidth]{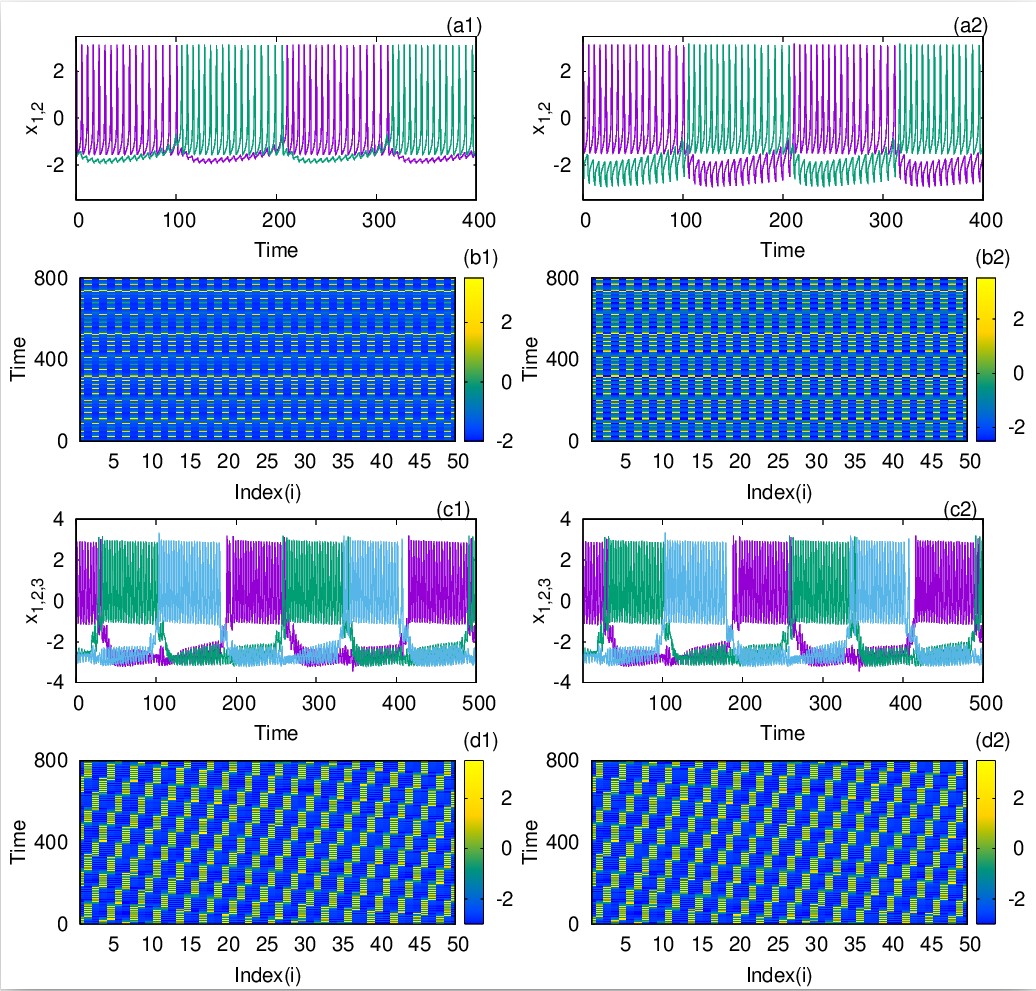}
\caption{Transfer of dynamical patterns from L2 to L1, in the 2-layer multiplex network with neurons on L2 coupled with inhibitory synaptic coupling and neurons in L1 uncoupled. Time Series of action potential and spatio-temporal plot are shown for different values of $p_2$ and $\lambda_2$: (a1,b1)first layer (a2,b2)second layer with $p_2=1$ and $\lambda_2=6$. (c1,d1) first layer and (c2,d2) second layer, with $p_2=2$ and $\lambda_2=10$. Here $\lambda_1=0$, $\epsilon=2$ and $N=50$.}
\label{f5}   
\end{figure}

\subsection{Dynamics of the multiplex network of neurons with excitatory and inhibitory synaptic couplings }
With the two layers of neurons multiplexed, we study how different emergent activity patterns of dynamics get selected across the layers as parameters are varied. We first consider the case where neurons of layer L1 are uncoupled, while those of layer L2 are coupled with inhibitory synaptic coupling, and both layers are coupled to each other via $i$ to $i$ connections with feedback coupling of strength $\epsilon$ as given in eqn~\ref{eq10}.
In this case, with $p_2=1$, $\lambda_2=6$ for L2, $\epsilon=1$, the patterns of synchronized oscillations that are anti-phase for adjacent nodes on second layer L2, get selected as such in the first layer L1 also. This is clear from Fig.~\ref{f5}(a1,a2) and Fig.~\ref{f5}(b1,b2), where the time series and spatio-temporal plots of both layers are given. Also for $p_2=2$ and $\lambda_2=10$ both layers show traveling wave-like oscillations (Fig.~\ref{f5}(c1,c2) and Fig.~\ref{f5}(d1,d2)). Thus, the emergent dynamics and the corresponding activity patterns get transferred from one layer to other layer when the layers are multiplexed.

\begin{figure}
\includegraphics[width=0.45\textwidth]{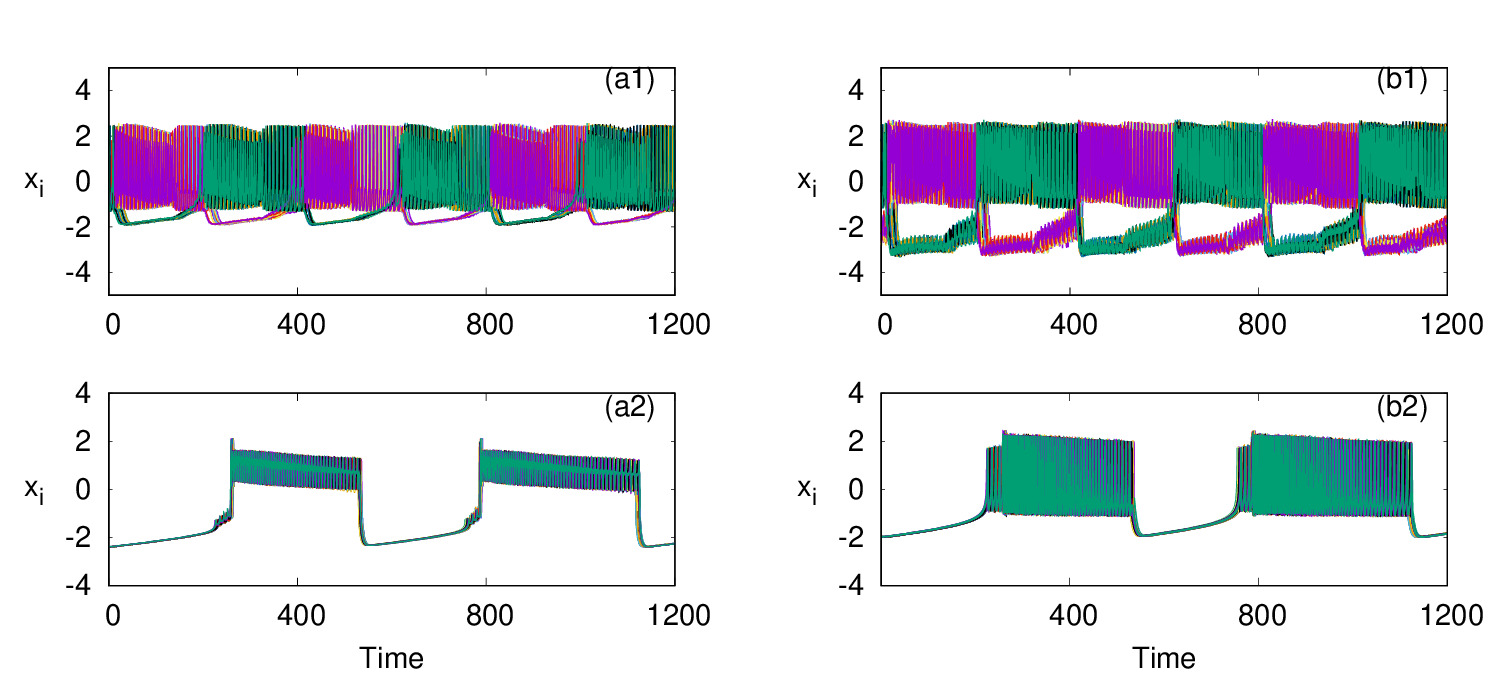}
\caption{Time series of the action potentials of multiplex HR neurons for both layers L1 (left panel)and L2 (right panel) for different values of $\lambda_1$ and $\lambda_2$: (a1,b1) $\lambda_1=0.1$ and $\lambda_2=1.0$, show anti-phase synchronized oscillations in two clusters, and (a2,b2) $\lambda_1=3$ and $\lambda_2=0.1$, show in-phase synchronized oscillations. Here $\epsilon=1$, $p_1=1$, $p_2=1$ and $N=50$. The pattern of the dynamics on the layer of larger intralayer coupling strength is selected across both layers.}
\label{f6}   
\end{figure}

Next, we consider the neurons of first layer L1 coupled with excitatory synaptic coupling, with neurons of L2 still coupled with inhibitory synaptic coupling, both with local couplings as $p_1=1$, and $p_2=1$.  With the interlayer coupling strength at $\epsilon=1$, for $\lambda_1=0.1$ and $\lambda_2=4$, we observe that both layers L1 and L2 exhibit anti-phase synchronized oscillations, with phase ordering which is shown in Fig.~\ref{f6}(a1,b1)respectively. When we set $\lambda_1=3.0$ and $\lambda_2=0.1$ , we observe  in-phase synchronized oscillations in both layers, which is shown in Fig.~\ref{f6}(a2,b2) respectively. 
Thus, we see that for strong inhibitory synaptic coupling strength, both layers show anti-phase synchronized oscillations in adjacent nodes, while for strong excitatory synaptic coupling, both layer show in-phase synchronized oscillations. 

\begin{figure}
\includegraphics[width=0.45\textwidth]{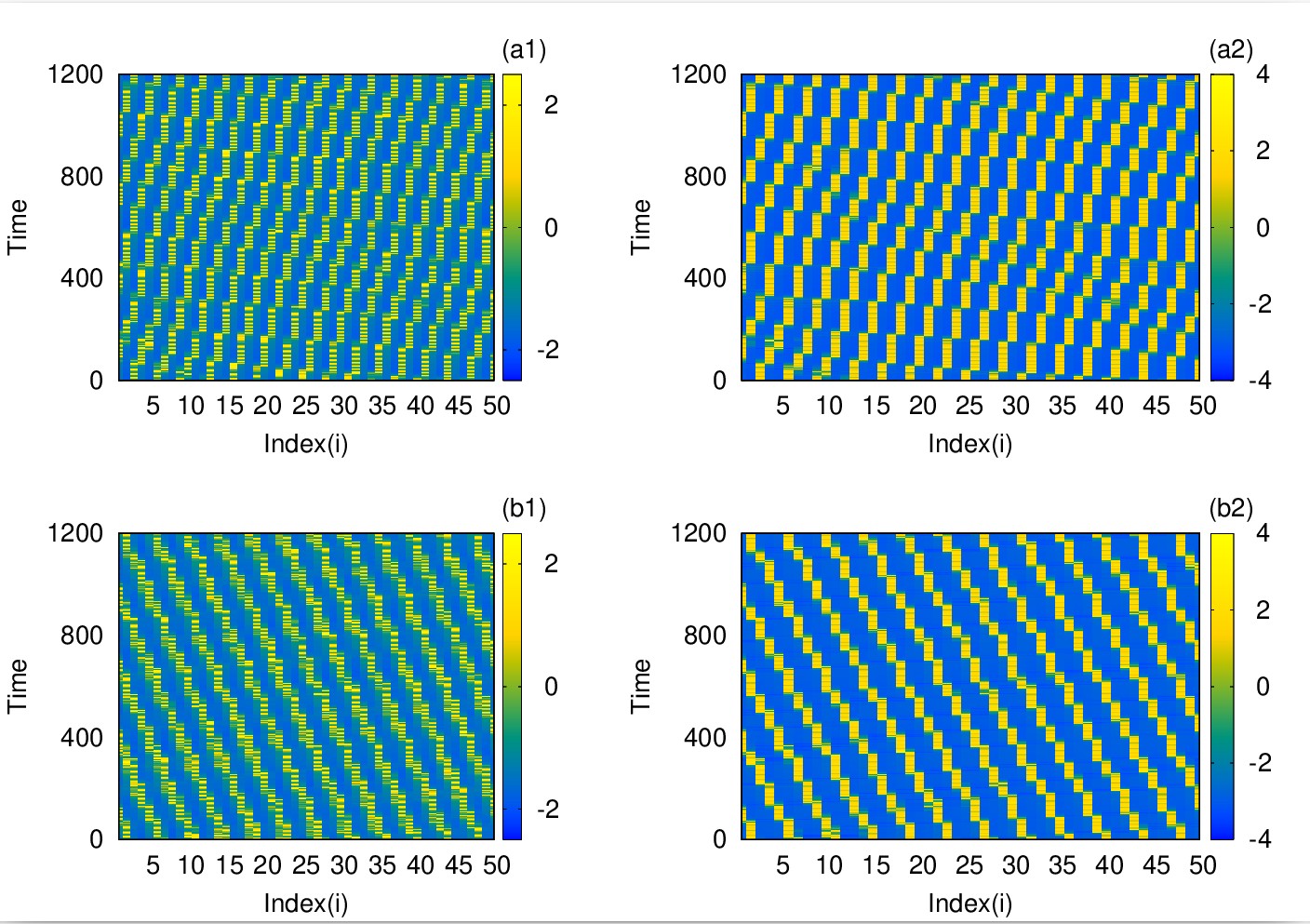}
\caption{Spatio-temporal plots of multiplex HR neurons for layer L1 (left panel)and L2 (right panel)for different values of $p_2$, and$\lambda_2$: (a1,b1) $p_2=2$ and $\lambda_2=6$, (a2,b2)$p_2=3$, and $\lambda_2=10$. Here $\lambda_1=0.1$, $\epsilon=1$ $p_1=10$, and $N=50$. The spatio-temporal dynamics of the layer with larger coupling strength gets selected in both the layers. However the nature of spikes and bursts are different in L1 and L2 due to difference in intralayer parameters.}
\label{f7}   
\end{figure}

Also, as couplings become nonlocal, with $p_2=2$ and $3$, both layers show phase shifted oscillations and spatio-temporal dynamics that are transferred from L2 to L1 for larger $\lambda_2$. We also observe that these states are selected by layer 1 for all values of $p_1$ up to $p_1=10$.  The spatio-temporal plots for $p_2=2$, and $\lambda_2=6$ shown in Fig.~\ref{f7}(a1,b1), and  for $p_2=3$ and $\lambda_2=10$, in Fig.~\ref{f7}(a2,b2),  indicate the transfer of dynamical patterns across the layers. However the nature of spikes and bursts in layers L1 and L2  differ due to difference in the parameter chosen.
So, the selection of the specific activity patterns on both layers depends on the relative intra-layer coupling strengths and follow the spatio-temporal dynamics of the layer with larger intra-layer coupling strength. This is further illustrated for other types of emergent dynamics below.

\begin{figure}
\includegraphics[width=0.45\textwidth]{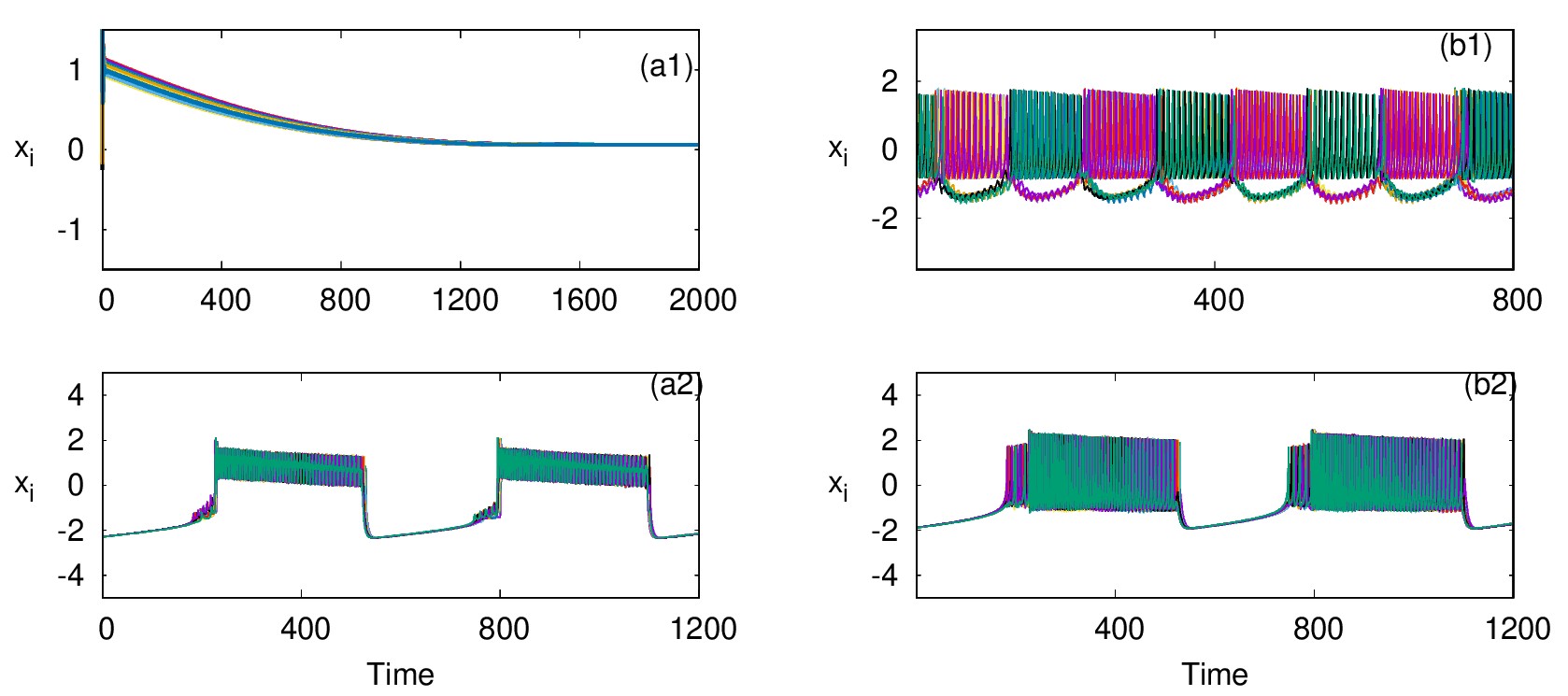}
\caption{Revival of activity in layer L1 due to multiplexing with L2 and transition from in phase to anti-phase in L2 as parameters are tuned. Time Series of action potentials for multiplex HR neurons in layer L1(left panel)and L2 (right panel)are plotted for different values of $\lambda_1$ and $\lambda_2$: (a1)at $\lambda_1=3$ $\epsilon=0$ neurons in layer L1 exhibit amplitude death,(b1) at $\lambda_2=0.3$ and $\epsilon=0$: neurons in layer L2 show anti-phase oscillations in two clusters: (a2,b2) $\lambda_2=0.3$ and $\epsilon=1$: observed revival of oscillations in L1 and transition from anti-phase to in-phase in L2. Here $\lambda_1=3$, $p_1=1$, $p_2=1$ and $N=50$.}
\label{f8}   
\end{figure}

\begin{figure}
\includegraphics[width=0.45\textwidth]{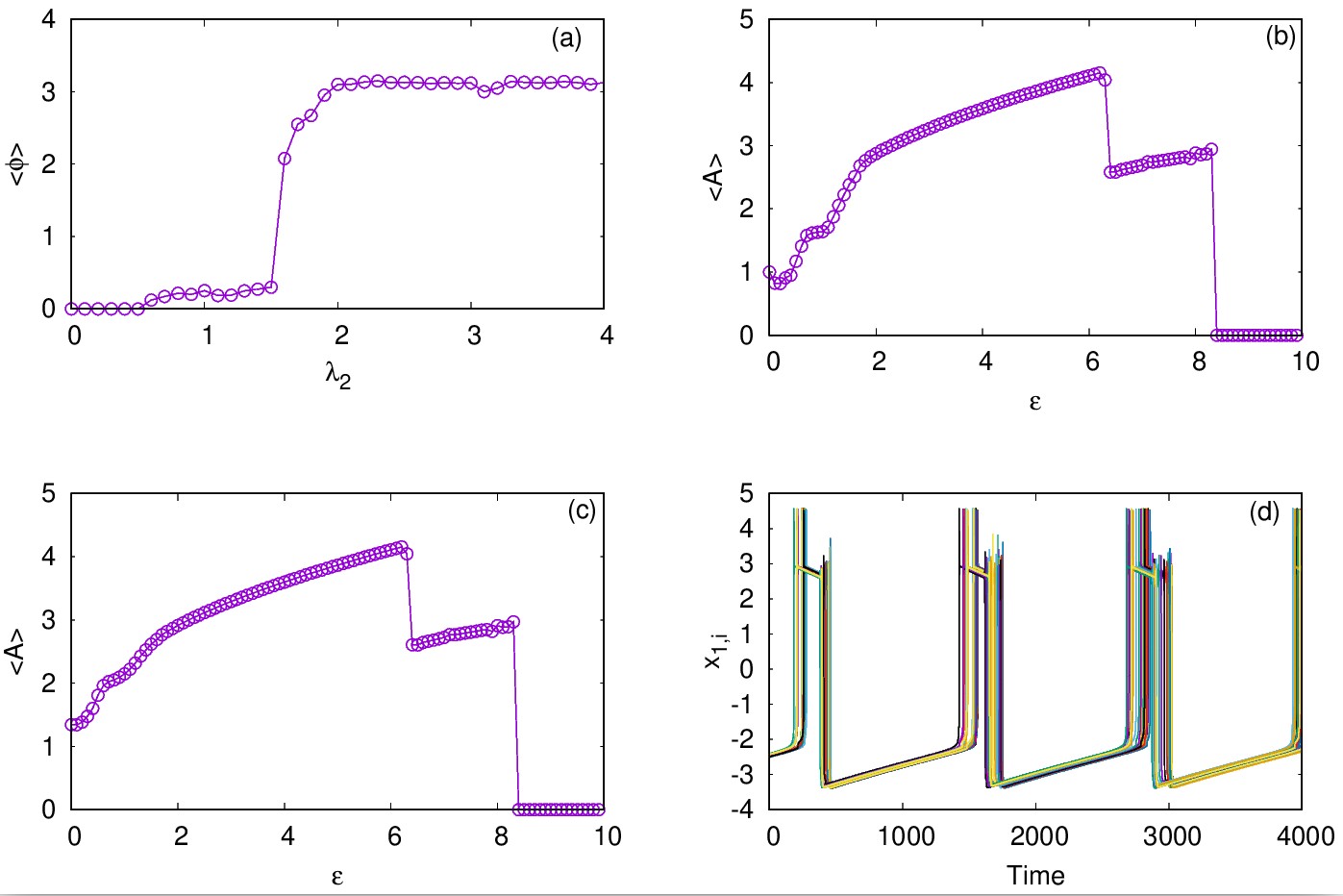}
\caption{(a) Transition from in-phase to anti-phase for synchronized activity in L2. Here average phase of neurons on second layer for varying coupling strength $\lambda_2$, is shown with neurons in L1 coupled locally with $p_1=1$ and $\lambda_1=3$.
Suppression of activity on increasing interlayer coupling strength. Average amplitudes of neurons on (b) L1  and (c) L2 , are shown with varying interlayer coupling strength $\epsilon$. The nature of spikes and bursts near the transition for $\epsilon=8.1$ is shown in (d).Here $\lambda_1=1$, $\lambda_2=1.0$, $p_1=1$, $p_2=1$ and $N=50$.}
\label{f9}   
\end{figure}

As reported earlier, when $\epsilon=0$, both L1 and L2 function as independent layers, and for higher synaptic coupling strength $\lambda_1=3$, layer L1 goes to amplitude death and at $ \lambda_2=0.3$, layer L2 show anti-phase synchronized oscillations in two clusters (Fig.~\ref{f8}(a1,b1)). But when both layers are multiplexed with $\epsilon=1$, we observe a revival of oscillations from death state on layer L1 and in-phase oscillation on layer L2, as shown in Fig.~\ref{f8}(a2,b2)respectively. Further we observe the activity pattern in L2 undergoes a transition from in-phase to anti-phase, as $\lambda_2$, is tuned. This transition from in-phase to anti-phase with increase in $\lambda_2$ in layer L2 is shown in Fig.~\ref{f9}(a), where the average phase difference is calculated as $<\phi>= \frac{1}{N}\sum_{i=1}^{N}(\phi_i-\phi_{i+1})$, with $\phi_i$ obtained for each neuron from  Eq.~\ref{eq2}. The inhibitory synaptic coupling in one layer can revive the oscillations from the suppressed state on the other layer. The variety of interesting activity patterns of spatio-temporal dynamics and their selection across layers happens at low to moderate values of interlayer coupling strengths. When the interlayer coupling strength $\epsilon$, is increased to say $\epsilon=10$, both layers settle to amplitude death states( Fig.~\ref{f9}(b,c)) and the time series near the transition point is as shown in Fig.~\ref{f9}(d).   Thus, the selection of activity patterns in both layers due to multiplexing depends on the nature and strengths of intralayer and interlayer couplings, and therefore, the coupling strengths and range of couplings can be tuned to select any desired pattern of activity.

\subsection{Dynamics of the multiplex network of neurons with electrical and synaptic coupling}
Now we consider the case where neurons in the first layer (L1) interact with each other with electrical coupling and those in the second layer (L2) interact through synaptic coupling. The dynamics of the multiplex network of neurons thus modelled is given as,

\begin{eqnarray}
\dot{x}_{i,1}&=&B_{i,1}+ \frac{\lambda_1}{2p_1}\sum_{k=i-p_1}^{i+p_1}(x_{k,1}-x_{i,1})+\epsilon  x_{i,2} \nonumber\\
\dot{y}_{i,1}&=&(a+\alpha)x_{i,1}^2-y_{i,1} \nonumber\\
\dot{z}_{i,1}&=&c(bx_{i,1}-z_{i,1}+e) \nonumber\\
\dot{x}_{i,2}&=&B_{i,2}+E\frac{\lambda_2}{2p_2}(V_s-x_{i,2})\sum_{k=i-p_2}^{i+p_2}\Gamma (x_{k,2})+\epsilon  x_{i,1} \nonumber\\
\dot{y}_{i,2}&=&(a+\alpha)x_{i,2}^2-y_{i,2} \nonumber\\
\dot{z}_{i,2}&=&c(bx_{i,2}-z_{i,2}+e),
\label{eq1}
\end{eqnarray}

Here we define a parameter E whose sign decides the nature of synaptic coupling, for $E= 1$ neurons in L2 are coupled with excitatory synaptic coupling, and for $E=-1$ second layer's neurons are coupled with inhibitory synaptic coupling.

\begin{figure}
\includegraphics[width=0.45\textwidth]{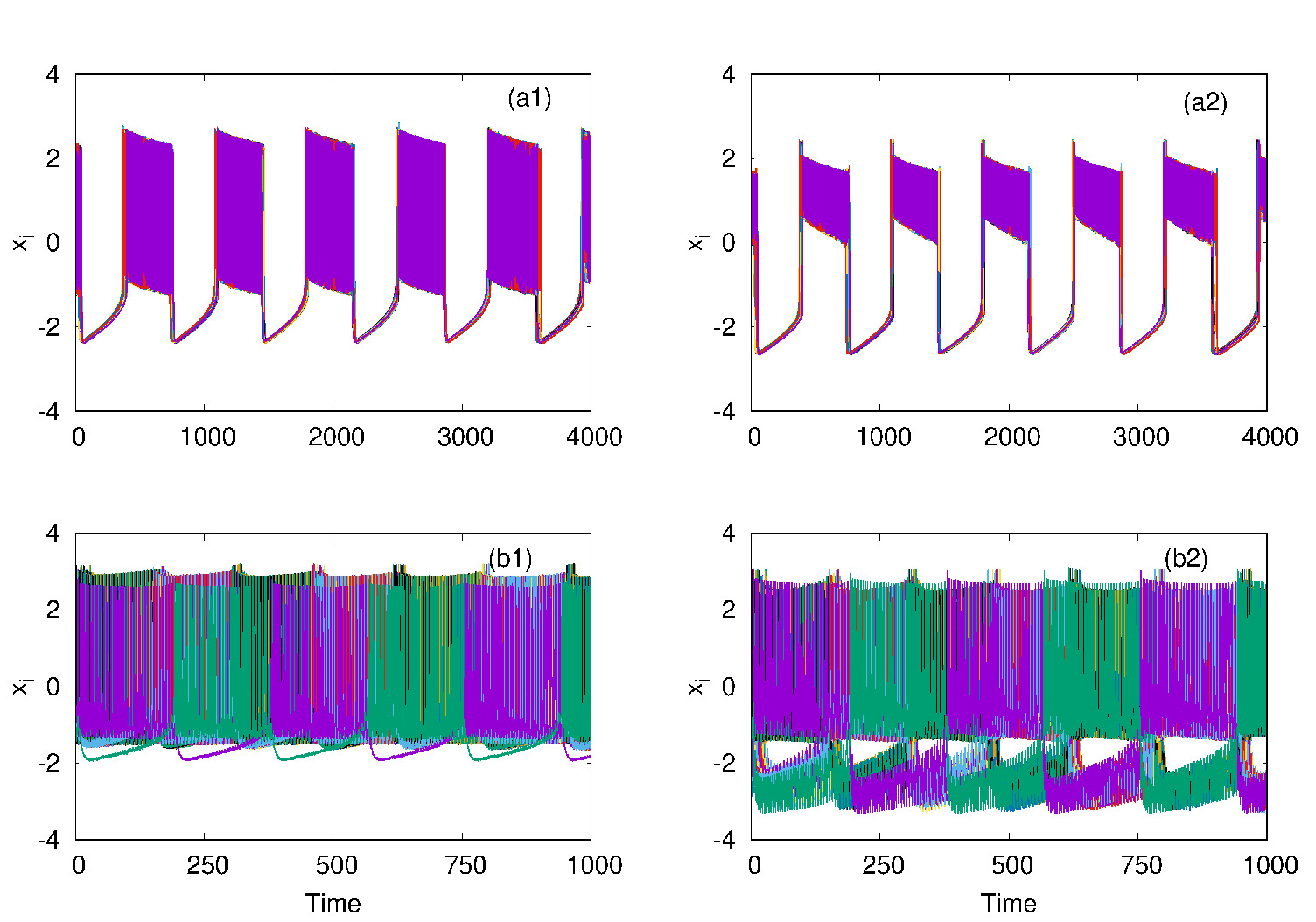}
\caption{ Time series of action potentials for multiplex HR neurons from layer L1 (left panel) and  L2 (right panel): (a1,b1) In-phase patterns for neurons of L1 coupled with electrical coupling and that of L2 coupled with excitatory synaptic coupling.(a2,b2) Anti-phase activity for neurons on L1 coupled with electrical coupling and that on L2 coupled with inhibitory synaptic coupling. The dynamics of L2 is transferred to L1 in both cases. $\lambda_1=0.5$, $\lambda_2=5$, $p_1=1$, $p_2=1$ and $N=50$.}
\label{f10}   
\end{figure}

\begin{figure}
\includegraphics[width=0.45\textwidth]{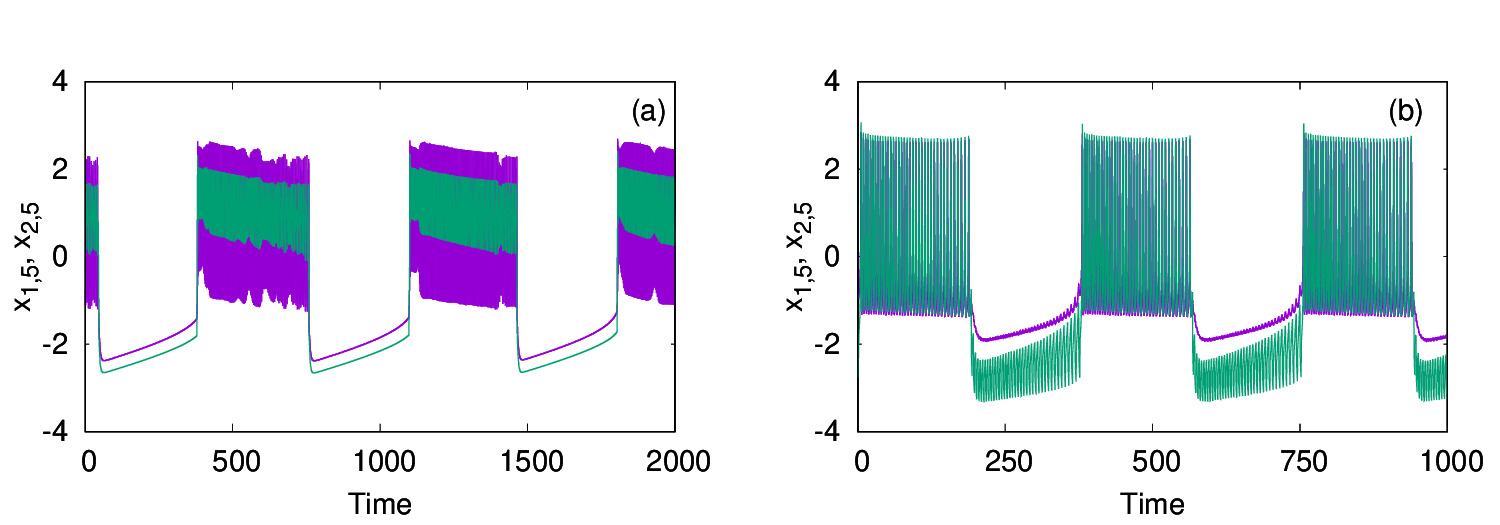}
\caption{In-phase synchronized dynamics between similar nodes on layers L1 and L2. Time series of node 5 from both layers L1 and L2 are plotted with (a) neurons of L1 coupled with electrical coupling and that of L2 coupled with excitatory synaptic coupling; (b) neurons of L1 coupled with electrical coupling and that of L2 coupled with inhibitory synaptic coupling. The other parameters are $\lambda_1=0.5$, $\lambda_2=5$, $p_1=1$, $p_2=1$ and $N=50$.}
\label{f11}   
\end{figure}

With E=1, and the excitatory coupling strength at $\lambda_2=5$, we observe that the coupled system shows in-phase synchronized oscillations, in both layers L1 and L2, as shown in Fig.~\ref{f10}(a1,b1) respectively.  Next, with E=-1, the coupled system shows anti-phase synchronized oscillations in both layers L1 and L2(Fig.~\ref{f10}(a2,b2)). Further, we also observe that along with the transfer of the emergent phenomena from one layer to another, the node of layer L1 shows in-phase synchronization with the same node of layer L2. To indicate this, we show the time series of the $5^{th}$ node of both layers L1 and L2, where layer L1 coupled with electrical coupling and L2 coupled with excitatory coupling in Fig.~\ref{f11}(a) and L2 coupled inhibitory synaptic coupling in Fig.~\ref{f11}(b) respectively.  Also, for strong electrical coupling strength ($\lambda_1$) and weak synaptic coupling ($\lambda_2$), (inhibitory or excitatory), we observe traveling wave patterns on both layers.

\section{Conclusion}

In this study, we report the selection of various activity patterns as the emergent spatio-temporal dynamics on a multiplex neuronal network of HR neurons where the nature of interaction in each layer can be different. This framework can thus model the plasticity and variability of connections among neurons which can exist as synaptic or electrical in nature with excitatory or inhibitory connections. By tuning the strengths of connections in each layer and across layers, the network can select various activity patterns and induce the pattern from one layer to the other. 

We first present the pattern of dynamics on the first layer L1, where neurons are coupled through excitatory synaptic couplings. By tuning the synaptic coupling strength, the coupled neurons can be in completely synchronized oscillations, while for strong synaptic coupling strength, the oscillations are suppressed to the state of amplitude death. The phenomenon of AD is observed for all values of $p_1$, corresponding to the local, nonlocal and global types of couplings. 
The second layer of neurons, coupled with inhibitory synaptic coupling, show anti-phase synchronized oscillations with amplification when the neurons are locally coupled, i.e., $p_2=1$. The anti-phase synchronized oscillations are interesting in two aspects. Firstly, the nature of oscillations are mixed-mode oscillations with enhanced frequency and amplitude with large amplitude spikes, and secondly, the phase relationship among them occurs in an orderly way, with alternate neurons being in phase and neighbouring ones being in anti-phase. Thus the whole network splits into two clusters, every odd node belonging to one cluster and every even node to the other cluster. For $p_2=2$ and $3$, we get traveling wave type of oscillations over the network. 

When the two layers are multiplexed, for sufficient inhibitory coupling strength, we observe mixed-mode synchronized oscillations that are phase-shifted get selected on both layers. In general, the selection of the specific pattern of activity on both layers can be controlled by tuning the relative intra-layer coupling strengths.

Also, multiplexing can revive the oscillations from the amplitude death state on the first layer by changing the inhibitory coupling strength on the second layer. We also report the transition from anti-phase to the in-phase type of mixed-mode oscillations, and vice versa that get selected as the excitatory and inhibitory coupling strengths are tuned to specific values. We repeat the study by increasing the size of the networks in both layers to 100 and 500 and find qualitatively similar results

With the nature of coupling among neurons in one layer L1 electrical, while the other layer L2 has neurons with synaptic connections, we observe in-phase synchronized activity in both layers when L2 has excitatory connections and anti-phase activity when it has inhibitory connections. We also find neurons at similar nodes in both layers are synchronous with in-phase oscillations. 
  
We note the variety of activity patterns presented here that occur for a collection of neurons forming a multiplex network, correspond to experimentally observed patterns of activity reported recently~\cite{croft}. Also, modulation of neuronal oscillation frequency is reported to occur during sensory information processing~\cite{Lee}. Thus the study provides a better understanding of the mechanism underlying such patterns known to occur in brain networks that incorporate multiplex network architecture naturally.~\cite{Frolov}.

\end{document}